\documentclass[a4paper,11pt,twoside]{article}
\usepackage[british]{babel}
\usepackage{amssymb,amsmath,amsfonts,verbatim,xkeyval,bm}
\usepackage{graphicx}
\usepackage{subfigure}
\usepackage{textcomp}
\def\naturali{\mathbb{N}}
\newcommand{\pluseq}{\hookleftarrow}

\def\build#1_#2^#3{\mathrel{
\mathop{\kern 0pt#1}\limits_{#2}^{#3}}}
\def\imunit{{\rm i}}
\newcommand{\csi}{\xi}
\newcommand{\Cscr}{{\cal C}}

\newcommand{\Hscr}{{\cal H}}
\newcommand{\Kscr}{{\cal K}}
\newcommand{\Lscr}{{\cal L}}
\newcommand{\Oscr}{{\cal O}}
\newcommand{\Pscr}{{\cal P}}
\newcommand{\Rscr}{{\cal R}}
\newcommand{\Sscr}{{\cal S}}
\newcommand{\Zscr}{{\cal Z}}
\def\poisson#1#2{\lbrace #1,#2 \rbrace}
\def\Lie{{\Lscr}}

\newtheorem{theorem}{Theorem}[section]

\newtheorem{proposition}[theorem]{Proposition}

\newtheorem{definition}[theorem]{Definition}

\title{\bf Trojan dynamics well approximated \\ by a new Hamiltonian
  normal form}

\author{{\bf Roc\'io Isabel P\'AEZ}\\ {\small Dipartimento di
    Matematica, Universit\`a degli Studi di Roma ``Tor
    Vergata'',}\\ {\small Via della Ricerca Scientifica 1, 00133--Roma
    (Italy).}\\ {\bf Ugo LOCATELLI}\\ {\small Dipartimento di
    Matematica, Universit\`a degli Studi di Roma ``Tor
    Vergata'',}\\ 
  {\small Via della Ricerca Scientifica 1, 00133--Roma
    (Italy).}\\ 
  {\small $\,$}\\ {\it Accepted for publication in Monthly Notices}\\ 
  {\it of the Royal Astronomical Society.} }

\date{}

\begin{document}
\maketitle

\bigskip

\markboth{R.I. P\'aez, U. Locatelli}{Trojans orbits
  reproduced by an integrable Hamiltonian}

\begin{abstract}
We revisit a classical perturbative approach to the Hamiltonian
related to the motions of Trojan bodies, in the framework of the
Planar Circular Restricted Three-Body Problem (PCRTBP), by introducing
a number of key new ideas in the formulation. In some sense, we adapt
the approach of~\cite{Garfinkel-77} to the context of the normal form
theory and its modern techniques.  First, we make use of Delaunay
variables for a physically accurate representation of the system.
Therefore, we introduce a novel manipulation of the variables so as to
respect the natural behavior of the model. We develop a normalization
procedure over the fast angle which exploits the fact that
singularities in this model are essentially related to the slow
angle. Thus, we produce a new normal form, i.e. an integrable
approximation to the Hamiltonian. We emphasize some practical examples
of the applicability of our normalizing scheme, e.g. the estimation of
the stable libration region.  Finally, we compare the level curves
produced by our normal form with surfaces of section provided by the
integration of the non--normalized Hamiltonian, with very good
agreement.  Further precision tests are also provided. In addition, we
give a step-by-step description of the algorithm, allowing for
extensions to more complicated models.
\end{abstract}

\maketitle

\section{Introduction}\label{sec:intro}

Series expansions in terms of small physical parameters are a common
way of dealing with dynamical models in Celestial Mechanics. One case
where this approach has been extensively used is the problem of Trojan
motion. The so-called Trojan stability problem, i.e. the study of the
long-term dynamics for massless particles in the neighborhood of the
equilateral Lagrangian points, has been studied both analytically and
numerically in several contexts\footnote{For an introduction to Trojan
  dynamics, see, e.g.,~\cite{Erdi-97}}.

From the numerical point of view, quite complete models (including
extra planets perturbations and/or 3-dimensional configurations) are
possible to consider. Such experiments are in general based on
accurate long integrations, done through high precision integrators. A
clear outcome is the determination of size and shape of the stable
domain around the equilibrium points.  Secondary resonances
  embedded within the domain of the 1:1 Mean Motion Resonance (MMR)
  are of particular importance in the stability issue.  For example,
in~\cite{Rob-et-al-05} and~\cite{Rob-Gab-06}, the importance of the
resonance web is highlighted for what concerns the process of
de-population of the Trojan domain, in the case of Jupiter's
asteroids.  In~\cite{Dvo-Lhot-Zhou-12}, the authors study the orbital
stability of the only discovered Trojan asteroid of the Earth,
i.e.~2010 TK7, and determine the stability region around L4 and L5 in
the Sun-Earth system.

Besides pure numerical investigations, several works have also
explored the applicability of analytical methods in the problem of
Trojan stability. In particular, the Trojan problem has served as a
classical model for testing methods based on the construction of a
so-called {\it Hamiltonian normal form} approach.

Semi-analytic methods share a common structure: each method is based
on an explicit algorithm, possible, as a rule, to translate into a
programming code, which computes the expansion of a suitable
\emph{normal form} that provides some special
solutions. Therefore, a normal form is a good \emph{local}
approximation to the complete Hamiltonian, used to reproduce the
orbits of particular objects.

Normal forms have been used in several cases for computing the
size and shape of the stability domain. From the point of view of
Nekhoroshev stability estimates
\cite{Nekhoroshev-1977},\cite{Nekhoroshev-1979}, examples of this
computation are in~\cite{Cel-Gio-91}, \cite{Gio-Sko-97},
\cite{Efthy-Sand-05}, \cite{Lhot-Efthy-Dvo-08}, \cite{Efthy-2013}. On
the other hand, in~\cite{Gab-Jor-Loc-05}, the stability estimation is
induced by applying the Kolmogorov-Arnold-Moser (KAM) theory. But all
these approaches, while introducing novel and promising features, tend
to fail when it comes to reproduce the numerical results about the
size and shape of the whole stability region. In general, they are
able to justify stability for just a small subset of Trojan orbits,
located relatively close to the equilibrium point.

In the present work, we develop a new normal form approach to the
classical PCRTBP by combining three main ideas. First, we make an
accurate choice of variables for the representation of the system.
Second, we explicitly take benefit of the existence of a slow and a
fast degree of freedom, dealing with them differently in our normal
form construction. Finally, we make use of Lie series techniques for
the averaging over the fast angle; in fact, as shown below, a key
element of our method is that such a treatment makes possible to
overcome the only true singularity that the Trojan motion contains,
namely the eventual \emph{close encounters} with the primary.

We motivate these ideas as follows.

Many of previous attemps describe the system in terms of cartesian
coordinates, which however do not capture properly the \emph{physical
  configuration} of the model. Considering canonical variables well
adapted to the system helps to improve the accuracy of the normal form
and the estimation of stability domains. To this end, in the present
paper, we adopt (and show the usefulness of) a modified version of
Delaunay-like coordinates introduced long ago in~\cite{Garfinkel-77}.

Regarding the other two ideas, an important point is that the only
real physical singularity in the 1:1 MMR region is due to possible
collisions~/~close encounters of the Trojan body with the primary. In
our setting of modified Delaunay-like variables (see
definition~\ref{eq:Delaunay-coord}), this singularity takes place at
$\lambda=0$, where $\lambda$ is the \emph{synodic mean longitude}
of the massless body. Thus, any polynomial expansion around the
equilibrium point, exhibits a bad convergence behavior for orbits
approaching this singularity. Nevertheless, in this work we show this
problem can be overcome in a rather simple way. This is based on the
following remarks.

On one hand, as noted already, an inspection of the main terms of the
Hamiltonian for a Trojan body, indicates that the motion is ruled by a
fast and a slow frenquency. In Delaunay variables, these two dynamics
are represented by two independent pairs of canonical coordinates.
However, the long term behavior of the orbits depends essentially only
on the slow degree of freedom which, in our variables, corresponds to
the canonical pair including the angle $\lambda$. In other words, an
appropriate normal form construction that aims to study the long term
dynamics involves averaging over the fast angle only.  This produces
an \emph{integrable} Hamiltonian of two degrees of freedom in which
the fast angle is ignorable.

On the other hand, such an averaging implies
that one has to solve a so-called homological equation in which
$\lambda$ plays no role. As shown in Section 2 below, due to this
property, one can retain in the Hamiltonian a complicated functional
dependence on $\lambda$, other than trigonometric or simply
polynomial. In our case this dependence turns to be of the
form of powers of the quantity $\beta(\lambda) = \frac{1}{\sqrt{2-2
    \cos \lambda}} $. Thus, using this technique allows to greatly
extend the convergence domain of the \emph{final} normal
form. 

Finally, we develop a new normalization algorithm for the computation
of the integrable approximation, by using the Lie series formalism,
adapting to a modern way the technique described
in~\cite{Garfinkel-77}. With this integrable normal form, we can
approximate Trojan orbits having either tadpole or horseshoe shapes,
even in cases where distances from the equilateral point become large
and where previous approaches tend to fail.

There are several different examples where such a normalizing scheme
can be applied. The most evident corresponds to the use of the scheme
for the estimation of the stability domain. Such a computation, in the
same direction as the references mentioned before, aim to estimate the
effects induced by the remainder $\Rscr$ (in later
Eq.~\ref{eq:H(R1,R2)}) of the normal form produced by the algorithm.
According to our results, our novel normalization may radically
improve the results regarding the size of the domain of stability
around the equilateral Lagrangian points, understimated so far. Other
examples inherent directly to the normal form can be mentioned. For
instance, in~\cite{Paez-Loc-14}, we used the normalized coordinates
for the design and optimization of maneuvers aiming to transfer a
spacecraft into the tadpole region. Moreover, the numerical
experiments described in the present work highlight that our method
allows to clearly differentiate the tadpole region from the horseshoe
region. In some cases, when the mass ratio between the two primary
bodies is very small and consequently also the chaotic regions are
small, such a computation gives a first order estimation of the stable
tadpole domain. On the other hand, this kind of perturbative approach
is motivating for problems of diverse nature. For instance,
in~\cite{Cec-Big-2013}, they make use of the Relegation algorithm,
that shares a similar structure with our normalizing scheme, to
compute some particular orbits around an irregularly shaped asteroid.

This paper is structured as follows: in Section~\ref{sec:expl_alg}, we
describe the expansion and normalization scheme, applied to the case
of the PCRTBP Hamiltonian, in the 1:1 MMR region; in
Section~\ref{sec:results_sect}, we provide different tests for a
suitable accuracy verification of the normal form, involving
comparisons with the numerical integration of the full
problem; in Section~\ref{sec:concl_future}, we summarize the
work and outline future applications of this method. Furthermore,
Appendix~\ref{sec:techn_things} formally presents the algorithm in
such a way that it can be adapted also to different models, for
example, the Elliptic Restricted Three Body Problem (ERTBP) or the
Restricted MultiPlanet Problem (RMPP), described in~\cite{Paez-Efthy-15}.

\section{Construction of the integrable approximation}\label{sec:expl_alg}

\subsection{Initial settings}\label{sbs:settings}

In heliocentric canonical variables $(\mathbf{p},\mathbf{r})$, the
Hamiltonian of the PCRTBP can be written as:
\begin{equation}\label{ham1}
H={p^2\over 2} -{G M\over r} - G m' \left({1\over\Delta}
-{\mathbf{r}\cdot\mathbf{r}'\over r'^3}\right)
\end{equation}
where $M$ and $m'$ are the masses of the larger and smaller primary,
respectively, $\mathbf{r}$ is the heliocentric position vector of the
test particle, $\mathbf{r'}$ the one corresponding to the second
primary, $p=\|\mathbf{p}\|$, $r=\|\mathbf{r}\|$, $r'=\|\mathbf{r'}\|$
and $\Delta = \| \mathbf{r} - \mathbf{r}' \|$.

In what follows, $a\,$, $e\,$, $M$ and $\varpi$ symbolize the major
semi-axis, the eccentricity, the mean anomaly and the longitude of the
perihelion (primed quantities correspond to the primary).  We set the
unit of length as $a'=1$, and the unit of time such that the mean
motion of the primary is equal to $n=1$. This implies $G(M+m')=1$.
The unit of mass is set so as to $G=1$.  Defining the mass parameter
as $\mu=m'$, the Hamiltonian in the above units takes the form:
\begin{equation}
H={p^2\over 2} -{1\over r} - \mu F~~,
\label{eq:ham2}
\end{equation}
where the so--called disturbing function $\mu F$ is such that
\begin{equation}
F=\left({1\over\Delta}
-{\mathbf{r}\cdot\mathbf{r}'\over r'^3}-{1\over r}\right)~~.
\label{eq:perturbing-term}
\end{equation}
Including the small keplerian correction $\mu/r$ in the disturbing
function allows to define action-angle variables with values
independent of the mass parameter $\mu$. Inspired
by~\cite{Paez-Efthy-15} and references therein, we introduce modified
Delaunay action-angle variables
\begin{equation}
\vcenter{\openup1\jot\halign{
 \hbox {\hfil $\displaystyle {#}$}
&\hbox {\hfil $\displaystyle {#}$\hfil}
&\hbox {$\displaystyle {#}$\hfil}
&\hbox {\hfil $\displaystyle {#}$}
&\hbox {\hfil $\displaystyle {#}$\hfil}
&\hbox {$\displaystyle {#}$\hfil}\cr
G &=& \sqrt{a(1-e^2)}\ ,
\qquad
&\lambda &=& M+\varpi-M^{\prime}-\varpi^{\prime}\ ,
\cr
\Gamma &=& \sqrt{a}\left(1-\sqrt{1-e^2}\right)\ ,
\qquad
&\ell &=& M\ .
\cr
}}
\label{eq:Delaunay-coord}
\end{equation}
According to the definitions of the orbital elements above,
$\lambda$ is the {\emph synodic} mean longitude.
In order to remove the \emph{fictitious} singularity of the Delaunay--like
variables when $e=0$ (i.e., for circular orbits), it is
convenient to introduce canonical coordinates
$(\rho,\csi,\lambda,\eta)$,  similar to Poincar\'e coordinates
for the {\em planar Keplerian problem}. Thus, let us define
\begin{equation}
\vcenter{\openup1\jot\halign{ \hbox {\hfil $\displaystyle {#}$} &\hbox
    {\hfil $\displaystyle {#}$\hfil} &\hbox {$\displaystyle {#}$\hfil}
    &\hbox {\hfil $\displaystyle {#}$} &\hbox {\hfil $\displaystyle
      {#}$\hfil} &\hbox {$\displaystyle {#}$\hfil}\cr \rho &= & G-1\ ,
    \qquad &\lambda &= & \lambda\ , \cr \csi &=
    &\sqrt{2\Gamma}\cos\ell\ , \qquad &\eta &=
    &\sqrt{2\Gamma}\sin\ell\ , \cr }}
\label{eq:Garfinkel-coord}
\end{equation}
where the values of $\rho$, $\csi$ and $\eta$ are significantly small
in a region surrounding the triangular Lagrangian points, for
instance, in the case of tadpole or horseshoe orbits (since $a\simeq
1$ and $e\gtrsim 0$).  Let us recall that those variables have been
introduced in~\cite{Garfinkel-77} to study the PCRTBP.

Before starting the construction of the normal form, it is necessary
to express Hamiltonian~\eqref{eq:ham2}, in particular the disturbing
function $F$ in~\eqref{eq:perturbing-term}, in terms of
Poincar\'e--Delaunay--like coordinates
$(\rho,\csi,\lambda,\eta)$. Here, we limit ourselves to sketch this
preliminary procedure\footnote{An exhaustive description of those
  expansions is out of the scopes of this publication. All the details
  are deferred to P\'aez, R.I., \emph{Ph.D. Thesis} (2015), that is in
  preparation and is available on request to the
  author.}. Essentially, as a first step, the terms $1/r$, $r^2$ and
$\mathbf{r}\cdot\mathbf{r}'$ (appearing in the Hamiltonian) must be
expanded with respect to orbital elements $a$, $e$, $M$ and $\lambda$,
following e.g. \S6 in \cite{MurrDermm-1999}. Afterwards, the
eccentricity $e$ is replaced by its expansion in power series of the
parameter $\sqrt{2\Gamma/G}$. By using the equations $a=(G+\Gamma)^2$,
$G=1+\rho$, $\Gamma=(\csi^2+\eta^2)/2$, we can then express $1/r$,
$r^2$ and $\mathbf{r}\cdot\mathbf{r}'$ as functions of the canonical
variables~\eqref{eq:Garfinkel-coord}.

As explained in the Introduction, the particular normalizing scheme we
use allows to keep a non polynomial nor trigonometric functional
dependence with $\lambda$. By making use of this idea, we keep powers
of $\beta(\lambda)=1/\sqrt{2-2\cos\lambda}$, in the expansions of the
term $1\over\Delta$ of $F$, which describes the main part of the
inverse of the distance from the primary (for values of the major
semi-axis $a\simeq 1$ and small eccentricity).  Thus, the Hamiltonian
ruling the motion of the third body takes the following
form \footnote{The first term in~\eqref{eq:initial-Ham-rewritten}
  comes from the expansion of the Keplerian part, i.e. the part of the
  Hamiltonian independent of $\mu$, in terms of $\csi$, $\eta$ and
  $\rho$, that corresponds to ${\cal K}=
  -\frac{1}{2(\rho+1+\frac{\csi^2+\eta^2}{2})} -\rho -1$.}
\begin{equation}
\vcenter{\openup1\jot\halign{ \hbox {\hfil $\displaystyle {#}$} &\hbox
    {\hfil $\displaystyle {#}$\hfil} &\hbox {$\displaystyle
      {#}$\hfil}\cr \Hscr(\rho,\csi,\lambda,\eta) &= & -\frac{1}{2}
    \sum_{j=0}^{\infty} (-1)^{j} (j+1)
    \left(\rho+\frac{\csi^2+\eta^2}{2} \right)^{j} \, -1 -\rho
       \,+\,\mu\big(1+\cos(\lambda)-\beta(\lambda)\big) \cr &+ \mu
    &\sum_{l=1}^{\infty}\,\sum_{{\scriptstyle{m_1+m_2}}\atop{\scriptstyle
        {+m_3=l}}} \ \sum_{{\scriptstyle{k_1+k_2\le
          l}}\atop{\scriptstyle {j\le 2l+1}}}
    b_{m_1,m_2,m_3,k_1,k_2,j}\,\rho^{m_1}\csi^{m_2}\eta^{m_3}
    \,\cos^{k_1}(\lambda)\sin^{k_2}(\lambda)\,\beta^j(\lambda)\ , \cr
}}
\label{eq:initial-Ham-rewritten}
\end{equation}
where $b_{m_1,m_2,m_3,k_1,k_2,j}$ are rational numbers.
Actually, we just produce a truncated expansion of
formula~\eqref{eq:initial-Ham-rewritten} up to a finite polynomial
degree in $\rho$, $\csi$ and $\eta$, by using {\tt Mathematica}.  By
inspection, we check that both properties $k_1+k_2\leq l$ and
$j\leq2l+1$ are satisfied in our initial expansion, and we just {\it
  conjecture} that they hold true also to any following polynomial
degree in $\rho$, $\csi$ and $\eta$ (being such a proof beyond of the
scopes of this work).

\subsection{Algorithm constructing the normal form averaged over
the fast angle}\label{sbs:average}

Let us focus on the first main terms of the Keplerian part:
\begin{equation}
-\frac{3}{2}+\frac{\csi^2+\eta^2}{2}
-\frac{3}{2}\left[\rho+\frac{\csi^2+\eta^2}{2}\right]^2 + \ldots\, =\,
-\frac{3}{2}+\Gamma-\frac{3}{2}(\rho+\Gamma)^2 + \ldots\ ,
\label{eq:quadr-part-Kepl-appr}
\end{equation}
where we refer to the dynamics of the new canonical
coordinates $(\csi,\eta)$ by retaining the old action--angle pair
$(\Gamma,\ell)$, that allow us to sketch our strategy in a
simpler way.  Indeed, the previous formula shows that the angular
velocities have different order of magnitude
\begin{equation}
\dot\lambda=\frac{\partial\,\Hscr}{\partial\rho}\simeq 0\ ,
\qquad
\dot\ell=\frac{\partial\,\Hscr}{\partial\Gamma}\simeq 1\ ,
\label{eq:slow-fast-ang-velocities}
\end{equation}
because $\mu\ll 1$ and the values of the actions $\rho$ and $\Gamma$
are small in a region surrounding the Lagrangian points. Thus,
$\lambda$ can be seen as a slow angle and $\ell$ as a fast angle. This
motivates to average the Hamiltonian over the fast angle (see, e.g.,\S~52 of \cite{Arnold-book-analyt-mech}), in order to focus mainly on
the secular evolution of the system. Therefore, we want remove (most
of) the terms depending on the fast angle $\ell$, that can be
conveniently done in the setting of the variables
$(\rho,\csi,\lambda,\eta)$, by performing a sequence of canonical
transformations. In the following, this strategy is translated in an
explicit algorithm using Lie series, by adapting to the present
context an approach that has been fruitfully applied to various
problems in Celestial Mechanics (for instance, for locating the
elliptic lower-dimensional tori in a rather realistic four-body
planetary model in~\cite{San-Loc-Gio-2011}, or studying the secular
behavior of the orbital elements of extrasolar planets
in~\cite{Lib-San-2013}). By applying such a procedure
  to our Hamiltonian model, we construct an average normal form,
  satisfying two important properties: (I)~it provides an accurate
  approximation of the starting model; (II)~it only depends on the
  actions and one of the angles and, therefore, it is integrable.

Our final normal form is produced by an algorithm dealing
with a sequence of Hamiltonians, whose the expansion can be
conveniently described after introducing the following
\begin{definition}
A generic function $g=g(\rho,\csi,\lambda,\eta)$ belongs to the class
$\Pscr_{l,s}\,$, if its expansion is of the type:
$$
\sum_{2m_1+m_2+m_3=l}
\ \sum_{{\scriptstyle{k_1+k_2\le l+4s-3}}\atop{\scriptstyle {j\le 2l+7s-6}}}
c_{m_1,m_2,m_3,k_1,k_2,j}\,\rho^{m_1}\csi^{m_2}\eta^{m_3}
\,(\cos\lambda)^{k_1}(\sin\lambda)^{k_2}\,\big(\beta(\lambda)\big)^j\ ,
$$
where $c_{m_1,m_2,m_3,k_1,k_2,j}$ are real coefficients.
\label{def-functions-classes}
\end{definition}

\noindent
Let $r_1$ and $r_2\,$ be two integer counters, running in the
intervals $[1\,,\,R_1]$ and $[0\,,\,R_2]$, respectively, being
$R_1\,,\,R_2\in\naturali$ fixed numbers. At each $(r_1,r_2)$--th step,
our algorithm introduces a new Hamiltonian $H^{(r_1,r_2)}$ such that
\begin{equation}
\vcenter{\openup1\jot\halign{
 \hbox {\hfil $\displaystyle {#}$}
&\hbox {\hfil $\displaystyle {#}$\hfil}
&\hbox {$\displaystyle {#}$\hfil}\cr
H^{(r_1,r_2)}(\rho,\csi,\lambda,\eta) &=
&\frac{\csi^2+\eta^2}{2}+
\sum_{l\ge 4}Z_l^{(0)}\Big(\rho,\frac{\csi^2+\eta^2}{2}\Big)
\cr
&+ & \sum_{s=1}^{r_1-1}\left(\sum_{l=0}^{R_2}
  \mu^s Z_l^{(s)}\Big(\rho,\frac{\csi^2+\eta^2}{2},\lambda\Big)
  +\sum_{l>R_2}\mu^{s}f_l^{(r_1,r_2;s)}(\rho,\csi,\lambda,\eta)\right)
\cr
&+ &\sum_{l=0}^{r_2}\mu^{r_1}
 Z_l^{(r_1)}\Big(\rho,\frac{\csi^2+\eta^2}{2},\lambda\Big)
\,+\,\sum_{l>r_2}\mu^{r_1}f_l^{(r_1,r_2;r_1)}(\rho,\csi,\lambda,\eta)
\cr
&+ &\sum_{s>r_1}\sum_{l\ge 0}\mu^sf_l^{(r_1,r_2;s)}(\rho,\csi,\lambda,\eta)\ ,
\cr
}}
\label{eq:H(r1,r2)}
\end{equation}
where $Z_l^{(0)}\in\Pscr_{l,0}$ $\forall\ l\ge 4$,
$Z_l^{(s)}\in\Pscr_{l,s}$ $\forall\ 0\le l\le R_2\,,\ 1\le s<r_1\,$,
$Z_l^{(r_1)}\in\Pscr_{l,r_1}$ $\forall\ 0\le l\le r_2\,$,
$f_l^{(r_1,r_2;r_1)}\in\Pscr_{l,r_1}$ $\forall\ l> r_2\,$,
$f_l^{(r_1,r_2;s)}\in\Pscr_{l,s}$ $\forall\ l>R_2\,,\ 1\le s<r_1\,$
and $\forall\ l\ge 0,\ s>r_1\,$. From ~\eqref{eq:H(r1,r2)}, we
emphasize that
\begin{itemize}
\item The splitting of the Hamiltonian in sub-functions belonging to
  different sets $\Pscr_{l,s}$ basically gather all the terms with the
  same order of magnitude $\mu^s$ and total degree $l/2$ (that can be
  semi-odd) in the actions $\rho$ and $\Gamma\,$. This is made in
  order to develop a normalization procedure, exploiting the existence
  of natural small parameters: $\mu$ and the values of the
  pair of actions $(\rho,\Gamma)$.
\item All the terms $Z_l^{(s)}$ and $f_l^{(r_1,r_2;s)}$ appearing in
  equation~\eqref{eq:H(r1,r2)} are made by expansions including a {\it
    finite} number of monomials of the type described in
  Definition~\ref{def-functions-classes}.
\item At the beginning of our algorithm, we can set $H^{(1,0)}=\Hscr$,
  because the expansion~\eqref{eq:initial-Ham-rewritten} of the
  initial Hamiltonian $\Hscr$ can be expressed as in
  equation~\eqref{eq:H(r1,r2)}.
\end{itemize}

Our algorithm requires just $R_1R_2$ normalization steps, which are
performed by constructing the finite sequence of Hamiltonians
\begin{displaymath}
H^{(1,0)}=\Hscr,\ H^{(1,1)},\ \ldots\,,\ H^{(1,R_2)},\ \ldots\,,
\ H^{(R_1,0)},\ H^{(R_1,1)},\ \ldots\,,\ H^{(R_2,R_1)}~~.
\end{displaymath}
They are defined so that $H^{(r_1+1,0)}=H^{(r_1,R_2)}$ $\forall\ 1\le
r_1<R_1$ and the $(r_1,r_2)$--th normalization step is performed by a
canonical transformation. This recursively introduces the new
Hamiltonian in such a way that
\begin{equation}
H^{(r_1,r_2)}=\exp\Big(\Lie_{\mu^{r_1}\chi_{r_2}^{(r_1)}}\Big)H^{(r_1,r_2-1)}\ ,
\label{eq:def-funzionale-H(r1,r2)}
\end{equation}
with a generating function
$\chi_{r_2}^{(r_1)}=\chi_{r_2}^{(r_1)}(\rho,\csi,\lambda,\eta)$ that
is determined by solving a so--called ``homological equation'', while
$\exp\big(\Lie_{\chi}\big)\,\cdot=\sum_{j\ge
  0}\frac{1}{j!}\Lie_{\chi}^j\,\cdot$ denotes nothing but the Lie
series operator. The Lie derivative $\Lie_{\chi} g=\poisson{g}{\chi}$
is such that $\poisson{\cdot}{\cdot}$ is the classical Poisson
bracket, with $g$ a generic function defined on the phase space and
$\chi$ any generating function (see, e.g.,~\cite{Giorgilli-2003.1}
for an introduction to canonical transformations expressed by Lie
series in the context of the Hamiltonian perturbation theory).  All
the recursive formulas, which determine the terms of type $Z$ and $f$
appearing in~\eqref{eq:H(r1,r2)}, are reported in
Appendix~\ref{sec:techn_things}. Let us stress that, after each
transformation, in the present subsection we do not change the name of
the canonical variables in order to simplify the notation. We
emphasize here that the new generating function introduced at the
generic $(r_1,r_2)$--th step, namely
$\mu^{r_1}\chi_{r_2}^{(r_1)}(\rho,\csi,\lambda,\eta)$, is determined
so as to remove from the main perturbing term\footnote{Let us recall
  that the size of $\mu^{s}f_{r_2}^{(r_1,r_2-1;s)}\in\Pscr_{r_2,s}$ is
  expected to decrease when the indexes $s$ or $r_2$ are increased,
  because the values of $\mu$, $\rho$ and $\sqrt{\csi^2+\eta^2}$ are
  assumed to be small} $\mu^{r_1}f_{r_2}^{(r_1,r_2-1;r_1)}$ its
subpart that is not in normal form.

We rewrite the final Hamiltonian in such a way to distinguish the
normal form part from the rest, as follows:
\begin{equation}
H^{(R_1,R_2)}(\rho,\csi,\lambda,\eta) =
\Zscr^{(R_1,R_2)}\big(\rho,(\csi^2+\eta^2)/2,\lambda\big)
+\Rscr^{(R_1,R_2)}(\rho,\csi,\lambda,\eta)\ ,
\label{eq:H(R1,R2)}
\end{equation}
where all the averaged terms of type $Z$ are gathered into the
integrable part $\Zscr^{(R_1,R_2)}$, while the others contribute to
the remainder $\Rscr^{(R_1,R_2)}$. Here, our algorithm is described at
a {\it purely formal} level in the sense that the problem of the
analytic convergence of the series on some domains is not
considered. However, it is natural to expect that our procedure
defines {\it diverging} series into the limit of $R_1,R_2\to\infty$,
because a non-integrable Hamiltonian cannot be transformed in an
integrable one, on any open domain.  In principle, the values of both
integer parameters $R_1$ and $R_2$ should be carefully chosen in
such a way to reduce the size of $\Rscr^{(R_1,R_2)}$ as much as
possible. In practice, we simply fixed the values of
$R_1$ and $R_2$ according to the computational resources, in order to
deal with the application described in the following
section~\ref{sec:results_sect}.

\subsection{Numerical computation of the flow induced by
the integrable approximation}\label{sbs:semi-analytical_integr-scheme}

Lie series induce canonical transformations in a Hamiltonian
framework; this fundamental feature allows us to design a numerical
integration method, by using both the normal form previously discussed
and the corresponding canonical coordinates. Let us denote with
$\big(\rho^{(r_1,r_2)},\csi^{(r_1,r_2)},\lambda^{(r_1,r_2)},\eta^{(r_1,r_2)}\big)$
the set of canonical coordinates related to the $(r_1,r_2)$--th
normalization step.  By applying the so--called 'exchange theorem'
(see~\cite{Giorgilli-2003.1}), we have that
\begin{equation}
H^{(r_1,r_2)}
\big(\rho^{{\scriptscriptstyle (r_1,r_2)}},\csi^{{\scriptscriptstyle (r_1,r_2)}},
\lambda^{{\scriptscriptstyle (r_1,r_2)}},\eta^{{\scriptscriptstyle (r_1,r_2)}}\big)
=H^{(r_1,r_2-1)}\Big(\varphi^{(r_1,r_2)}
\big(\rho^{{\scriptscriptstyle (r_1,r_2)}},\csi^{{\scriptscriptstyle (r_1,r_2)}},
\lambda^{{\scriptscriptstyle (r_1,r_2)}},\eta^{{\scriptscriptstyle (r_1,r_2)}}\big)\Big)\ ,
\label{eq:exchange-theorem}
\end{equation}
where the variables related to the previous step, i.e.
$(\rho^{{\scriptscriptstyle (r_1,r_2-1)}},\csi^{{\scriptscriptstyle (r_1,r_2-1)}},
\lambda^{{\scriptscriptstyle (r_1,r_2-1)}},\eta^{{\scriptscriptstyle (r_1,r_2-1)}})$,
are given as
\begin{equation}
\varphi^{(r_1,r_2)}
\big(\rho^{(r_1,r_2)},\csi^{(r_1,r_2)},\lambda^{(r_1,r_2)},\eta^{(r_1,r_2)}\big)=
\exp\Big(\Lie_{\mu^{r_1}\chi_{r_2}^{(r_1)}}\Big)
\big(\rho^{(r_1,r_2)},\csi^{(r_1,r_2)},\lambda^{(r_1,r_2)},\eta^{(r_1,r_2)}\big)\ .
\label{eq:coord-change-(r1,r2)}
\end{equation}
According to this, Lie series must be applied separatedly to each
variable, in order to properly define all the coordinates for the
canonical transformation $\varphi^{(r_1,r_2)}$. Thus, the whole
normalization procedure is described by the canonical transformation
\begin{equation}
\Cscr^{(R_1,R_2)}=\varphi^{(1,1)}\circ\ldots\circ\varphi^{(1,R_2)}\circ\ldots
\circ\varphi^{(R_1,1)}\ldots\circ\varphi^{(R_1,R_2)} \ .
\label{eq:def-total-canonical-transf}
\end{equation}
Such a composition of all the intermediate changes of variables can be
used for providing the following semi-analytical scheme to integrate
the equations of motion:
\begin{equation}
\vcenter{\openup1\jot\halign{
  \hbox to 34 ex{\hfil $\displaystyle {#}$\hfil}
&\hbox to 11 ex{\hfil $\displaystyle {#}$\hfil}
&\hbox to 34 ex{\hfil $\displaystyle {#}$\hfil}\cr
\left(\rho^{(0,0)}(0),\csi^{(0,0)}(0),
\lambda^{(0,0)}(0),\eta^{(0,0)}(0)\right)
&\build{\longrightarrow}_{}^{{{\scriptstyle
\big(\Cscr^{(R_1,R_2)}\big)^{-1}}
\atop \phantom{0}}}
&\left(\rho^{(R_1,R_2)}(0),\csi^{(R_1,R_2)}(0),
\lambda^{(R_1,R_2)}(0),\eta^{(R_1,R_2)}(0)\right)
\cr
& &\big\downarrow \build{\Phi_{\Zscr^{(R_1,R_2)}}^{t}}_{}^{}
\cr
\left(\rho^{(0,0)}(t),\csi^{(0,0)}(t),
\lambda^{(0,0)}(t),\eta^{(0,0)}(t)\right)
&\build{\longleftarrow}_{}^{{{\scriptstyle \Cscr^{(R_1,R_2)}} \atop \phantom{0}}}
&\left(\rho^{(R_1,R_2)}(t),\csi^{(R_1,R_2)}(t),
\lambda^{(R_1,R_2)}(t),\eta^{(R_1,R_2)}(t)\right)
\cr
}}
\qquad\qquad\ ,
\label{semi-analytical_scheme}
\end{equation}
where $\Phi_{\Kscr}^{t}$ denotes the flow induced on the canonical
coordinates by the generic Hamiltonian $\Kscr$ for an interval of time
equal to $t$. Let us emphasize that the above integration scheme
provides just an {\it approximate} solution. From an ideal point of
view (i.e., if all the expansions were performed without errors and
truncations), formula~\eqref{semi-analytical_scheme} would be exact if
the normal form part $\Zscr^{(R_1,R_2)}$ would correspond to the
complete Hamiltonian $H^{(R_1,R_2)}$; moreover, $\Zscr^{(R_1,R_2)}$ is
{\it integrable} and its flow is easy to compute\footnote{In order to
  explicitly describe the solutions of the equation of motions for the
  normal form $\Zscr^{(R_1,R_2)}$, it is convenient to introduce the
  temporary action--angle variables
  $\big(\Gamma^{(R_1,R_2)},\ell^{(R_1,R_2)}\big)$ such that
  $\csi^{(R_1,R_2)}=\sqrt{2\Gamma^{(R_1,R_2)}}\cos\ell^{(R_1,R_2)}$
  and
  $\eta^{(R_1,R_2)}=\sqrt{2\Gamma^{(R_1,R_2)}}\sin\ell^{(R_1,R_2)}$,
  where $\Gamma^{(R_1,R_2)}$ is a constant of motion for the normal
  form
  $\Zscr^{(R_1,R_2)}=\Zscr^{(R_1,R_2)}\big(\rho^{(R_1,R_2)},\Gamma^{(R_1,R_2)},\lambda^{(R_1,R_2)}\big)$. By
  considering $\Gamma^{(R_1,R_2)}$ as a fixed parameter and using the
  standard quadrature method for conservative systems with $1$~d.o.f.,
  one can compute $\rho^{(R_1,R_2)}(t)$ and $\lambda^{(R_1,R_2)}(t)$
  at any time $t\,$. The same can be done for the evolution of
  $\ell^{(R_1,R_2)}(t)$, by evaluating the integral corresponding to
  the differential equation
  ${\dot\ell}^{(R_1,R_2)}=\frac{\partial\,\Zscr^{(R_1,R_2)}}{\partial\Gamma^{(R_1,R_2)}}\,$. For
  practical purposes, the application of the classical quadrature
  method can be replaced by any numerical integrator that is precise
  enough. Finally, the values of $\csi^{(R_1,R_2)}(t)$ and
  $\eta^{(R_1,R_2)}(t)$ can be directly calculated from those of the
  corresponding action--angle variables, that are
  $\Gamma^{(R_1,R_2)}(t)$ and $\ell^{(R_1,R_2)}(t)$.}, reasons why
using $\Zscr^{(R_1,R_2)}$ becomes valuable. According to
equation~\eqref{eq:H(R1,R2)}, the approximate solution provided by the
scheme~\eqref{semi-analytical_scheme} is as more accurate as smaller
the perturbing part $\Rscr^{(R_1,R_2)}$ is with respect to
$\Zscr^{(R_1,R_2)}$.

A key point concerns the structure of our expansions. If we
{\it truncate} the r.h.s. of equation~\eqref{eq:H(r1,r2)} up to a
finite order of magnitude in $\mu$ and a fixed total degree $l/2$ in
the actions $\rho$ and $\Gamma=(\csi^2+\eta^2)/2$, since each term of
type $Z$ and $f$ is included in a corresponding class of functions
$\Pscr_{l,s}\,$, the truncated expansion of $H^{(r_1,r_2)}$ just
include a {\it finite} number of monomials. The same applies also to
the truncated expansions of both the final Hamiltonian $H^{(R_1,R_2)}$
and the canonical transformation $\Cscr^{(R_1,R_2)}$.  Therefore,
after introducing the truncation rules, the normalization algorithm
require a {\it finite} total number of operations, to compute the
elements necessary to implement the whole integration
scheme~\eqref{semi-analytical_scheme}. Thus, it can be translated in a
programming code.

In order to concretely apply our semi-analytical scheme, by using {\tt
  Mathematica}, we compute the truncations up to the terms
$\Oscr(\mu^3)$ and to the fifth total degree with respect to the
square roots of the actions $\rho$ and $(\csi^2+\eta^2)/2$
(i.e. $R_1=3$, $R_2=5$, in order to obtain $\Zscr^{(3,5)}\simeq
H^{(3,5)}$, $\Cscr^{(3,5)}$ and its inverse).  At the end of its
execution, that code writes, on an external {\it file}, {\bf C}
functions which are able to compute the changes of coordinates induced
by $\Cscr^{(3,5)}$ and by its inverse. Moreover, it provides also
another external {\it file}, where the integrable equations of motion
related to the Hamiltonian $\Zscr^{(3,5)}$ are written according to
the syntax of the {\tt Taylor}\footnote{{\tt Taylor} is an automatic
  translator producing a {\bf C} function acting as a time-stepper
  specific for a given ordinary differential equation (by means of the
  Taylor method). It is publicly available at the following website:
  {\tt http://www.maia.ub.es/{$\sim$}angel/soft.html}} software
package.  A basic use of the {\tt Linux} {\it shell scripting} allowed
us to automatically complement all these parts of the computational
procedure, so as to produce a new {\bf C} program as an output. This
final code is able to numerically integrate the equations of motion
via the scheme~\eqref{semi-analytical_scheme}, where the flow
$\Phi_{\Zscr^{(3,5)}}^{t}$ is computed by using the Taylor method
(based on the automatic differentiation technique,
see~\cite{Jor-Zou-2005} and~\cite{Barrio-et-al-2012}).  The truncation
rules are fixed in such a way to execute all the computations in a
reasonable amount of CPU-time.

\section{Accuracy control for the normal form}\label{sec:results_sect}
\nobreak An averaged model approximating a certain problem is powerful
to the extent that it reproduces the main features of the original
system, (in this case the Planar Circular Restricted Three Body
problem). It is widely proved that PCRTBP is barely a very simplistic
representation for the Trojan motion, but its basic dynamics allows us
to test in a simple way the normalization method. Furthermore,
according to the nature of the method, much more complex (i.e.,
non-planar, eccentric, including additional planets) models can be
treated without any substantial change in the averaging scheme.

When seen from the most commonly used reference system (i.e. a
\emph{synodic} rotating frame with origin in the barycenter), motions
around the equilateral Lagrangian points are represented by the
so-called \emph{tadpole} or \emph{horseshoe} orbits (see \S3.9 of
\cite{MurrDermm-1999}). In our set of variables, these orbits are
characterized by large variations of the angle $\lambda$. So, as
starting, a suitable averaged approximation has to be able to
reproduce these variations correctly. In particular, this should apply
even for the challenging case of bodies whose orbits lay very close to
the border of the stability domain. As second goal, the averaged
Hamiltonian should be able also to distinguish between tadpole orbits
(around just one equilateral equilibrium point) and horseshoe orbits
(around both points). In Section~\ref{sec:expl_alg}, we have provided
an integrable averaged Hamiltonian $\Zscr^{(R_1,R_2)}$
(see~\ref{eq:H(R1,R2)}), that explicitly describes the behavior of the
degree of freedom related to the canonical pair of variables
($\rho$,$\lambda$).  In order to compare the original problem with our
averaged version, we develop two different tests.

\subsection{Numerical surfaces of section vs. semi-analytical level curves}\label{sbs:graph_test}
\nobreak The first test consists of a graphical comparison between the
orbits provided by the complete Hamiltonian and those provided by
normal form $\Zscr^{(R_1,R_2)}$. In both cases, the initial conditions
are fictitious, but each set is derived from the catalogued
position of a real generating Trojan body. Since we are just
interested in the behavior of the slow degree of freedom, for the
complete problem (originally 4D), we just retain the evolution of the
variables $\lambda$ and $\rho$ by means of isoenergetic surfaces of
section, which gives a 2D representation. In the two cases presented
below, the \emph{generating bodies} are 2010 TK$_7$, Trojan asteroid
of the Sun-Earth system and the asteroid 1872 Helenos of the
Sun-Jupiter system.

From a catalogue, we obtain the coordinates (rotated to the plane
  of primaries) of each generating body for a certain epoch and we
convert them to our canonical coordinates
$(\rho_{gb},\lambda_{gb},\csi_{gb},\eta_{gb})$. This initial condition
provides the Jacobi constant $C_{J_{gb}}$ for the body, after which it
is possible to produce an isoenergetic surface of section. We generate
a set of 10 new orbits by keeping fixed the initial values for the
variables $\rho=\rho_{gb}$ and $\eta=\eta_{gb}$, scanning a range of
values for $\lambda$ and obtaining $\csi$ in such a way that
$C_J(\rho,\lambda,\csi,\eta) = C_{J_{gb}}$ (isoenergetic orbits).
These initial conditions are numerically integrated for a short time,
up to the variables accomplish the relation $M(\csi,\eta) = 0$, where
$M$ corresponds to the mean anomaly. We call this new set
$\Sscr_{gb}$, and it generates both the surface of section and the
level curves.

\subsubsection{Computation of surfaces of section}\label{sss:surf_sec}

Starting from Newtonian formulation for the PCRTBP, we derive the
equations of motion for the third body in cartesian coordinates in the
heliocentric system\footnote{Although cartesian coordinates are not
  canonical conjugated variables, they provided the simplest
  closed form for these equations of motion, and for this
  reason they are still widely used for numerical experiments.}
\begin{equation}\label{eq:eqs-mot-complete-crtbp}
\vcenter{\openup1\jot\halign{
  \hbox {\hfil $\displaystyle {#}$}
&\hbox {\hfil $\displaystyle {#}$\hfil}
&\hbox {$\displaystyle {#}$\hfil}
\qquad
&\hbox {\hfil $\displaystyle {#}$\hfil}
&\hbox {\hfil $\displaystyle {#}$\hfil}
&\hbox {\hfil $\displaystyle {#}$\hfil}\cr
\dot{x} & = & v_x\ ,
&\dot{v}_x & =
& - \frac{(1-\mu)x}{\sqrt{x^2+y^2}}
- \mu \left(\frac{x- x_P}{\sqrt{(x-x_P)^2+(y-y_P)^2}}
+ \frac{x_P}{\sqrt{x_P^2+y_P^2}} \right)\ ,
\cr
\dot{y} & = & v_y\ ,
&\dot{v}_y & = & - \frac{(1-\mu)y}{\sqrt{x^2+y^2}}
-\mu \left( \frac{y- y_P}{\sqrt{(x-x_P)^2+(y-y_P)^2}}
+ \frac{y_P}{\sqrt{x_P^2+y_P^2}} \right)\ .
\cr
}}
\end{equation}
where $x$, $y$, $v_x$, $v_y$ correspond to cartesian positions and
velocities of the massless body, and $x_P$, $y_P$ give the
instantaneous position of the planet.  We translate initial conditions
of the set $\Sscr_{gb}$ to cartesian coordinates and we integrate them
with a Runge-Kutta $7\>$-$\>8^{\>{\rm th}}$ order integrator, along
1500 periods of primaries, with time-step equal to $2\pi/100$.  During this
integration we collect the points contained in the pericentric surface
of section, which in our case is represented by the condition
\begin{equation}\label{eq:peric-surf-sect}
M (\csi,\eta) = 0 \quad \mathrm{or\:equivalently} \quad \eta = 0 ~~,
\end{equation}
and provide about 1500 points per orbit. The output data is again
translated to Delaunay variables and suitable to be compared with the
results of the averaged Hamiltonian.

\subsubsection{Computation of level curves}\label{sss:lev_curv}

Through Hamilton's equations, we can
easily derive the equations of motion for the canonical variables
$\rho$ and $\lambda$,
\begin{equation}\label{eq:haml-eqs}
\dot{\rho} = -\frac{\partial \Zscr^{(R_1,R_2)}}{\partial \lambda} \qquad \dot{\lambda}
= \frac{\partial \Zscr^{(R_1,R_2)}}{\partial \rho} ~~.
\end{equation}
Since $\Zscr^{(R_1,R_2)}$ is given by a series expansion of the
variables $\rho$, $\Gamma$, $\cos{\lambda}$, $\sin{\lambda}$,
$\beta(\lambda)$ and the small parameter $\mu$, the equations of
motion inherit the same structure.  Through a suitable storing of the
coefficients of these series and management of the equations of motion
implemented in {\tt Mathematica} (as explained in last paragraph of
Subsection~\ref{sbs:semi-analytical_integr-scheme}), we compute the
evolution of the orbits according to the averaged Hamiltonian.  Every
initial condition of the set $\Sscr_{gb}$ is first converted to
normalized variables and then integrated up to collecting about 2000
points, keeping the relative energy error of the integration smaller
than $10^{-12}$.  Such a numerical integration is an efficient way to
compute the {\it level curves} for the integrable normal form
$\Zscr^{(R_1,R_2)}$ corresponding to the values
$\Gamma=(\csi^2+\eta^2)/2$ and
$\Zscr^{(R_1,R_2)}(\rho,\Gamma,\lambda)$ (in normalized variables).

We complement every point of a level curve with the values
  $\csi=\sqrt{2\Gamma}$ and $\eta=0$ (equivalent to $M=0$). Let us
note here that the condition $M=0$ in the normalized coordinates does
not correspond exactly to the surface of section $M=0$ in the original
variables. However, since the change of coordinates
$\Cscr^{(R_1,R_2)}$ (that gives the values of the non--normalized
variables in the scheme~\eqref{semi-analytical_scheme}) is, by
construction, a near-to-identity canonical transformation, we assume
that the conditions for each surface of section do not differ too much.
Finally, via $\Cscr^{(R_1,R_2)}$, we back-transform all the points of
a level curve in the original variables and we graphically compare
them with the corresponding numerical surface of section.

\subsubsection{Examples and results}\label{sss:examples_results}

\begin{figure}
  \includegraphics[width=.35\textwidth,angle=270]{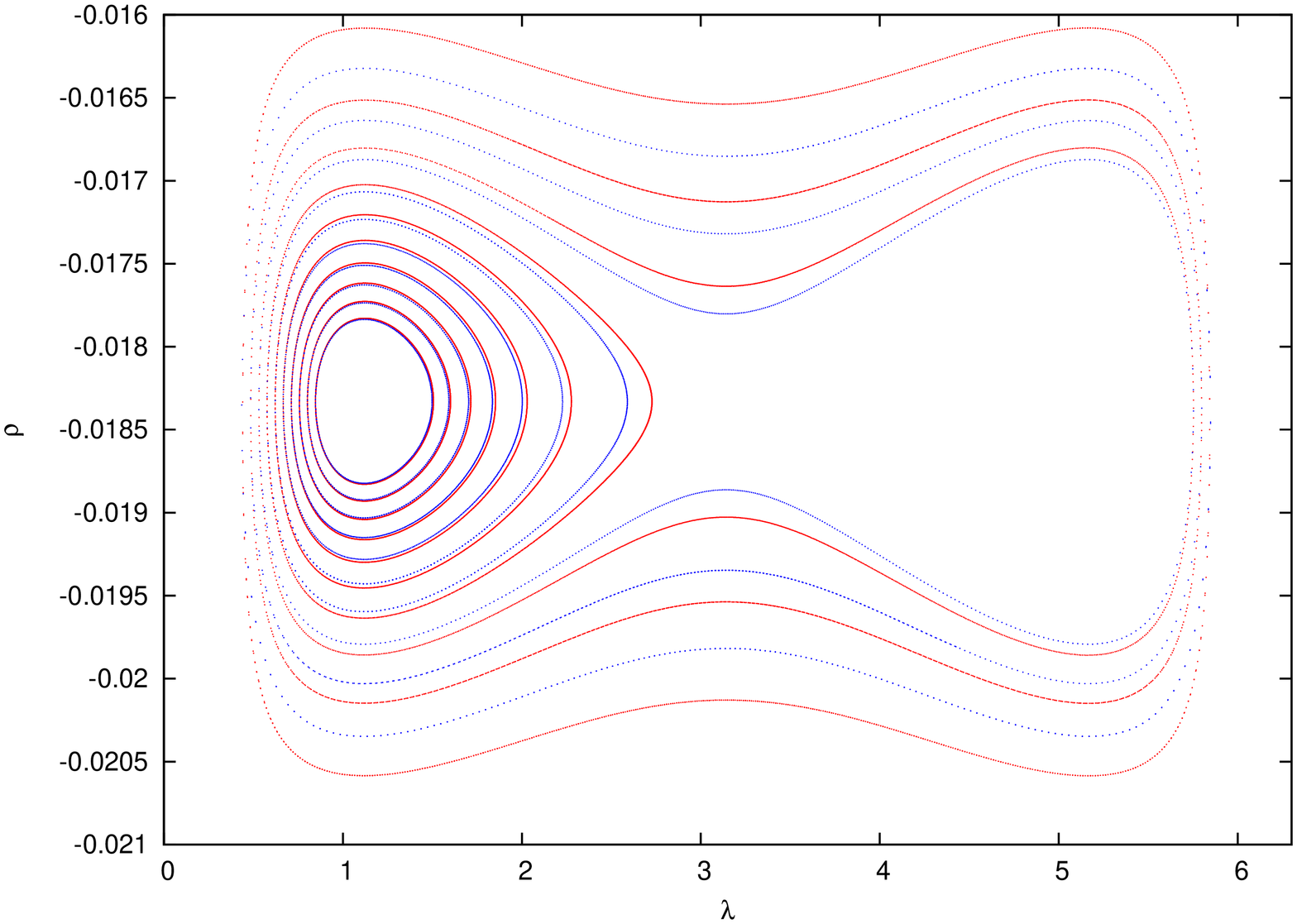}
  \includegraphics[width=.35\textwidth,angle=270]{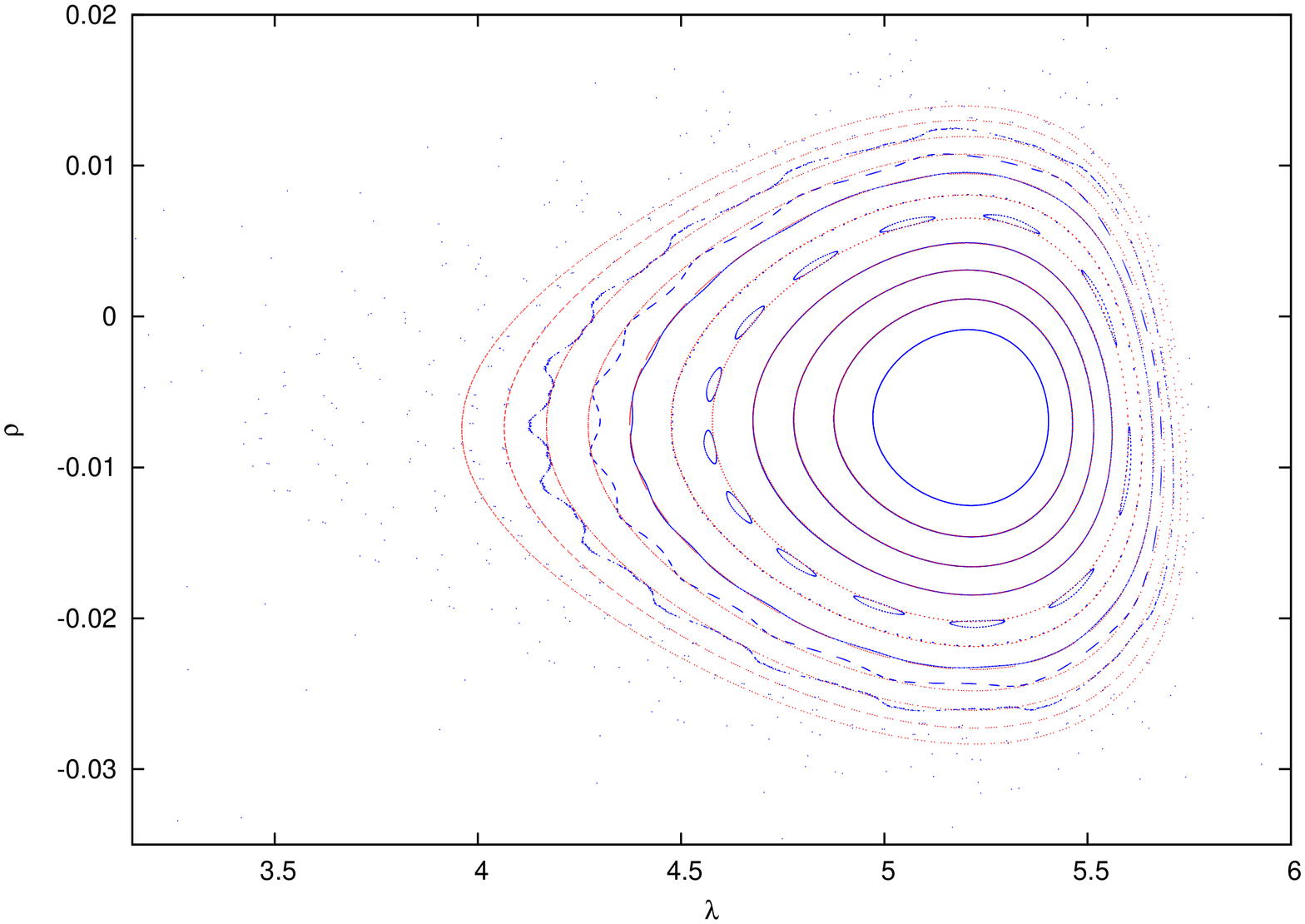}
  \caption{Comparison between the level curves produced by the
    averaged Hamiltonian in red (light gray) and the points of the
    surface of section for the complete problem in blue (dark gray),
    for the Sun-Earth problem (left panel) and Sun-Jupiter problem
    (right panel). In the Sun-Earth case, the generating body is the
    Earth Trojan 2010 TK$_7$. In the Sun-Jupiter case, the generating
    body is Trojan asteroid 1872 Helenos. See text for more
    details.}
  \label{fig:com_surf}
\end{figure}

We choose two systems with very different values of the mass parameter
for a better contrast in the results of the test.  The first case is
provided by the PCRTBP approach to the Sun-Earth system, which is
defined by a mass parameter $\mu=0.30003 \times 10^{-5}$.  The
generating body chosen for this system is the Earth only Trojan 2010
TK$_7$, for which we obtain its coordinates from Jet Propulsion
Laboratories JPL Ephemerides Service\footnote{{\tt
    http://ssd.jpl.nasa.gov/?ephemerides}}, at epoch 2456987.5 JD
(2014-Nov-26). We translate them to our set of canonical variables,
resulting $\rho_{TK7} = -1.8401447 \times 10^{-2}$, $\lambda_{TK7} =
3.5736334$, $\eta_{TK7} = 0.1152511$ and $\csi_{TK7} = -0.1530054\,$.
For the second case, we choose the Sun-Jupiter system, defined by the
mass parameter $\mu=0.953855\times10^{-3}$. The generating body in
this case is the Trojan asteroid 1872 Helenos, which belongs to
  the Trojan camp around L5 in such a system. The set of initial
conditions for Helenos were obtained from Bowell Catalogue\footnote{
  {\tt http://www.naic.edu/$\sim$nolan/astorb.html}}, at 2452600.5~JD
(2002-Oct-22), and after being translated to canonical Delaunay
variables, they read $\rho_{1872} = -0.3836735\times10^{-2}$,
$\lambda_{1872} = 5.6716748$, $\eta_{1872} = -0.0154266$ and
$\csi_{1872} = -0.1104177\,$.

Figure~\ref{fig:com_surf} shows the comparison between the surface of
section and the level curves computed for Sun-Earth system (left
panel), and Sun-Jupiter system (right panel). For both cases, the
points of the surface of section are represented in blue (dark gray)
and the curves produced with the averaged Hamiltonian are in red
(light gray). For the case of Sun-Earth system the agreement between
the two representations is excellent. According to the milestones we
define at the beginning of this section, the averaged Hamiltonian
reproduces accurately the large variations of $\lambda$. In
particular, it is perfectly able to distinguish between orbits
belonging to the tadpole or to the horseshoe region. In the case
Sun-Jupiter system, for which the mass parameter value is 3 orders of
magnitude larger, there is a substantial presence of
chaos. Nevertheless, the averaged Hamiltonian is able to \emph{locate}
any tadpole orbit provided by the complete Hamiltonian, even in cases
when the motion is trapped into secondary resonances, and its validity
is also good for orbits close to the border of the stable region.

\subsection{Computation of quasi-actions}\label{sbs:comp_act}

So far in the literature, one the most successful attemps to construct
a normal form for testing the stability of Trojans in the context of
the Hamiltonian formalism is discussed
in~\cite{Gab-Jor-Loc-05}. However, of the 34 initial conditions they
used for the system Sun-Jupiter, 4 presented orbits that were highly
chaotic after being translated according their rotations into the
planar CRTBP, while the Kolmogorov normalization algorithm defined in
that work did not work properly for other 7 cases. Here, we revisit
the fictitious initial conditions they considered for these latter
seven asteroids (1868~Thersites, 1872~Helenos, 2146~Stentor,
2207~Antenor, 2363~Cebriones, 2674~Pandarus and 2759~Idomeneus). Since
they conform a set of coordinates that either lay very close to
the border of stability or show an anomalous behavior (with respect to
the expected tadpole orbit), they provide a natural harder test in a
more quantitative way.

\begin{figure}
  \centering
  \includegraphics[width=.5\textwidth,angle=270]{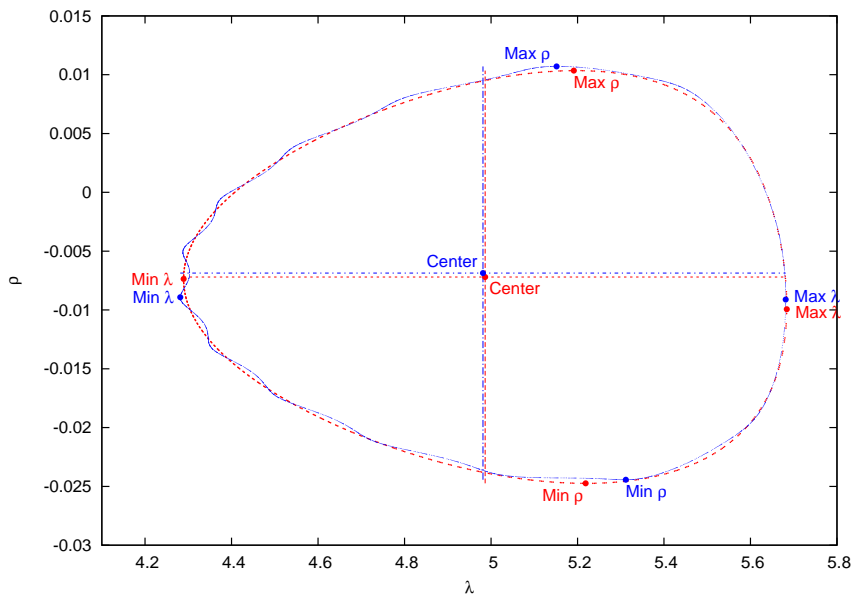}
  \caption{Location of the maximum and miminum values for the
    coordinates $\rho$ and $\lambda$ in the integration of the
    fictitious initial condition associated to 1872 Helenos, around
    L5. Red (light gray) points correspond to the averaged level curve
    and blue (dark gray) points to the numerical surface of section}
  \label{fig:strg_guy}
\end{figure}

In the previous subsection, we show that the evolution of
the orbits given by $\Zscr^{(R_1,R_2)}$ emulates correctly in many
cases the evolution under the original non-normalized Hamiltonian, by
means of graphical comparisons between level curves and surfaces of
section. This normal form $\Zscr^{(R_1,R_2)}$ contains two different
actions or integrals of motion. One, obtained by construction and
through the normalization, is given by $\Gamma$. The other one, not
explicitly obtained, is due to the fact that, after the reduction of
$\Gamma$, the normal form bear just 1 d.o.f., i.e. it is
integrable. This second constant of motion is associated with the area
enclosed by the level curve computed under the integration of
$\Zscr^{(R_1,R_2)}$, and therefore, provide another quantity to be
checked.  In order to do so, the computation of the orbits is done as
explained in
subsections~\ref{sss:surf_sec}--\ref{sss:examples_results}, but for
just one initial condition.  After integrating both curves, we compute
the maximum and minimum values for the two variables $\rho$ and
$\lambda$ reached during the integration. In Fig.~\ref{fig:strg_guy},
we show the positions of those quantities, over the surfaces of
section defined in subsection~\ref{sbs:graph_test}. With those values,
first we obtain the positions of the centers for the two orbits
through
\begin{equation}\label{eq:center-avrg}
\mathrm{C}_{\mathrm{avrg}} = \left(
\mathrm{C}_{(\lambda,\mathrm{avrg})},
\mathrm{C}_{(\rho,\mathrm{avrg})} \right)= \left(
\frac{\left(\mathrm{Max}\,\lambda_{\mathrm{avrg}} -
  \mathrm{Min}\,\lambda_{\mathrm{avrg}} \right)}{2}, \frac{ \left(
  \mathrm{Max}\,\rho_{\mathrm{avrg}} -
  \mathrm{Min}\,\rho_{\mathrm{avrg}} \right) }{2} \right) ~~,
\end{equation}
and
\begin{equation}\label{eq:center-num}
\mathrm{C}_{\mathrm{num}} = \left(
\mathrm{C}_{(\lambda,\mathrm{num})}, \mathrm{C}_{(\rho,\mathrm{num})}
\right)= \left( \frac{\left(\mathrm{Max}\,\lambda_{\mathrm{num}} -
  \mathrm{Min}\,\lambda_{\mathrm{num}} \right)}{2}, \frac{ \left(
  \mathrm{Max}\,\rho_{\mathrm{num}} -
  \mathrm{Min}\,\rho_{\mathrm{num}} \right) }{2} \right) ~~.
\end{equation}
The dispersion between centers, which shows how much displacement
there is between the orbits, is evaluated by putting
\begin{equation}\label{eq:disp}
\delta \mathrm{C} = \frac{\left(\mathrm{C}_{(\rho,\mathrm{num})} -
  \mathrm{C}_{(\rho,\mathrm{avrg})}
  \right)}{\mathrm{C}_{(\rho,\mathrm{num})}} +
\frac{\left(\mathrm{C}_{(\lambda,\mathrm{num})} -
  \mathrm{C}_{(\lambda,\mathrm{avrg})}
  \right)}{\mathrm{C}_{(\lambda,\mathrm{num})}} ~~.
\end{equation}
Furthermore, we obtain the value of the enclosed areas for each curve,
as follows: we first refer all the points composing the curve, to the
already computed center, by introducing the quantities $\delta
\lambda=\lambda-C_{\lambda}$ and $\delta\rho=\rho-C_{\rho}\,$;
therefore, we obtain the distance to the center $\mathrm{d} =
\sqrt{\delta \rho^2 + \delta \lambda^2}$ and the angle $\theta$ with
respect to the horizontal line ($ \theta = {\rm atan2}
(\delta\rho\,,\,\delta \lambda)$). Re-ordering the points by
increasing value for the angle $\theta$, we compute the area contained
in the triangle generated by two consecutive points and the center.
The sum along all the triangles represents the contained area within
each curve, $A_{\mathrm{num}}$ for the complete system and
$A_{\mathrm{avrg}}$ for average Hamiltonian flow. For a further
comparison, we compute also the relative difference between the areas
$\delta
A/A_{\mathrm{num}}=|A_{\mathrm{num}}-A_{\mathrm{avrg}}|/A_{\mathrm{num}}$,
 and the displacement of the centers of the orbits with respect to
  the position of the equilateral Lagrangian point they librate around
  ($\rho = 0$, $\lambda_{L4}=\pi/3$ for $L_4$ and
  $\lambda_{L5}=5\pi/3$ for $L_5$).
\begin{table}
	\centering
	\caption{Summary of the results for the quantities defining each
averaged and numerical orbit}
	\label{tab:table_data}
\begin{tabular}{|cccccc|}
\hline
Asteroid & $A_{\mathrm{num}}$ & $\delta A/A_{\mathrm{num}}$ & $\delta \mathrm{C}$ & $\mathrm{C}_{(\rho,\mathrm{num})}$ & $\mathrm{C}_{(\lambda,\mathrm{num})}-\lambda_{L4,L5}$ \\
\hline
\hline
1868 & $2.03\times 10^{-2}$ &  $\, 3.21\times 10^{-3}$& $\, 6.95\times 10^{-3}$& $-1.08\times 10^{-2}$ & $-0.163$ \\
\hline
1872 & $3.75\times 10^{-2}$ &  $\, 1.39\times 10^{-3}$& $\, 5.14\times 10^{-2}$ & $-6.86\times 10^{-3}$ & $-0.235$ \\
\hline
2146 & $1.67\times 10^{-2}$ &  $\, 1.25\times 10^{-1}$& $\, 3.71\times 10^{-2}$ & $-1.94\times 10^{-1}$ & $-0.530$ \\
\hline
2207 & $2.31\times 10^{-2}$ &  $\, 6.59\times 10^{-3}$& $\, 7.50\times 10^{-3}$& $-1.31\times 10^{-2}$ & $-0.196$ \\
\hline
2674 & $3.56\times 10^{-3}$ &  $\, 1.51\times 10^{-2}$& $\, 3.61\times 10^{-3}$& $-1.43\times 10^{-2}$ & $-0.077$ \\
\hline
2759 & $2.67\times 10^{-2}$ &  $\, 1.29\times 10^{-2}$& $\, 1.04\times 10^{-2}$& $-1.63\times 10^{-2}$ & $-0.232$ \\
\hline
\end{tabular}

\end{table}

\noindent

For 6 of the 7 cases mentioned above, we are able to obtain averaged
orbits that reflects the behavior of the numerical integrations. In
Table~\ref{tab:table_data}, where we report the results for the
previously defined quantities, we show that the averaged areas clearly
match their associated numerical areas, with a relative error smaller
than 2\%, except for one case (asteroid 2146 Stentor), for which the
error is about 13\%. This may be due to the fact that 2146 Stentor
presents the largest displacement with respect to the corresponding
Lagrangian point, in a quite anomalous orbit. For the rest of the
asteroids, the position of the orbits in the surface of section turns
to be very close to that generated by the equivalent averaged level
curve, both centered at a triangular Lagrangian point. On the other
hand, in Table 1 we do not present data for the highly
inclined\footnote{According to the Bowell Catalogue at~2452600.5 JD}
($39^{\circ}$) asteroid 2363~Cebriones of our sample. For this
asteroid, our normal form fails to provide an accurate orbit, using
the initial conditions provided in~\cite{Gab-Jor-Loc-05}. However, we
find that the numerical orbit generated by 2363~Cebriones presents a
very peculiar angular excursion (in $\lambda$) with respect to the
Lagrangian point. This failure of the normal form could be generated
in the initial condition by a non-consistent rotation to the plane of
the primaries, in the original work.

\section{Summary and perspectives}\label{sec:concl_future}

In this paper we present a novel normalization scheme, that provides
an integrable approximation of the dynamics of a Trojan asteroid
Hamiltonian in the framework of the Planar Circular Restricted
Three-Body Problem (PCRTBP). This new algorithm is based on three
co-related points: the introduction of a set of variables which
respects the physical configuration of the system; the existence of
two degrees of freedom with well distinguishable roles, one
corresponding to a fast motion and another corresponding to a slow
motion, that allows us to fruitfully average over the fast angle;
finally, the analytic singularity of this model exclusively related
with the slow angle.

These three concepts motivates a new way to deal with the initial
expansion of the Hamiltonian, as it is necessary for the normalization
procedure. The slow angle $\lambda$ does not affect the solution of
the homological equation, determined in order to remove the dependence
on the fast angle. Thus, we are able to carefully approximate level
curves that represent tadpole and horseshoe orbits by keeping, in the
expansions, a non polynomial dependence just with respect to
$\lambda$.

In order to examinate the accuracy of the normal form produced, we
develop some tests. We study numerically integrated surfaces of
section, along the flow of the complete Hamiltonian, and we
contrast them with the level curves provided by our integrable normal
form, with very good agreement. Furthermore, we estimate some
quasi-integrals of motion (by computation of enclosed areas), which
also show excellent agreement with those corresponding to the
numerically computed surfaces of section.  On the whole, this novel
approach for producing a new normal form results in a very promising
approximation of the global behavior of the Trojans motion.

From a rather theoretical point of view, we think that our normalizing
scheme can be complemented with a scheme of estimates, so as to make
the remainder exponentially small on a suitable open domain. If our
algorithm is joint with a Nekhoroshev-like approach, it should ensure
that the eventual diffusion is effectively bounded (i.e., for
intervals of time comparable with the age of our Solar system), for a
set of initial conditions. We expect such a set to be significantly
larger than those considered in the works already existing in the
literature. This expectation is due to the fact that our method offers
a wide coverage of the Trojans orbits. For the same reason, we think
that our integrable approximation can be used in many cases to
successfully start the Kolmogorov normalization algorithm. While the
corresponding solution is valid for any time, the construction of an
invariant KAM torus is an extremely local procedure, because it must
be adapted to the orbit of each Trojan body to be studied.  In this
context, we think that our algorithm can be efficiently used
jointly with a KAM-like approach in those cases for which the
previous implementation of the Kolmogorov normalization algorithm
failed.  Let us remark that such a new application of the KAM theory
would require to preliminarly construct the action--angle coordinates
also for the slow degree of freedom, before starting the final
normalization procedure. As an alternative strategy, a different
formulation of the KAM theorem that is not strictly based on
action--angle variables could be used ~\cite{Lla-Gon-Jor-Vil-2005}.

More practically, in our opinion the most exciting point is that our
method is suitable to be translated to more complex models, without
requiring essential changes. Since the PCRTBP corresponds to a very
simplistic representation for the Trojans domain, we are presently
working to extend the normalization to a Hamiltonian that also
considers the eccentricity of the primary. The first preliminary
results show that our method can be used so as to locate the main
secondary resonances within the 1:1 MMR region. In particular, we find
a good agreement with other purely numerical indicators, when the mass
ratio between the primaries is small. We plan to include these and
other results in a future publication. Furthermore, we think that our
present and future contributions will help to fill that still existing
gap between the semi-analytic studies and the more complete numerical
experiments of the stability region.

\section*{Acknowledgements}

The authors would like to thank C.~Efthymiopoulos for his advice and
constant support. We are indebted also with C.~Sim\'o who suggested to
reconsider the article~\cite{Garfinkel-77}, and with the anonymous
referee, whose contribution helped improving the original
manuscript. During this work, R.I.P. was supported by the Astronet-II
Marie Curie Training Network (PITN-GA-2011-289240), while U.L. was
partially supported also by the research program ``Teorie geometriche
e analitiche dei sistemi Hamiltoniani in dimensioni finite e
infinite'', PRIN 2010JJ4KPA\_009, financed by MIUR.








\appendix
\section{Details about the formal algorithm constructing the normal form}\label{sec:techn_things}
\nobreak
As discussed in subsection~\ref{sbs:average}, the normalization
algorithm defines a sequence of Hamiltonians. This is done by an
iterative procedure; let us describe the basic step which introduces
$H^{(r_1,r_2)}$ starting from $H^{(r_1,r_2-1)}$ when both the values
of the indexes $r_1$ and $r_2$ are positive. We assume that the
expansions of $H^{(r_1,r_2-1)}$ is such that
\begin{equation}
\vcenter{\openup1\jot\halign{
 \hbox {\hfil $\displaystyle {#}$}
&\hbox {\hfil $\displaystyle {#}$\hfil}
&\hbox {$\displaystyle {#}$\hfil}\cr
H^{(r_1,r_2-1)}(\rho,\csi,\lambda,\eta) 
&= &\frac{\csi^2+\eta^2}{2}+
\sum_{l\ge 4}Z_l^{(0)}\Big(\rho,\frac{\csi^2+\eta^2}{2}\Big)
\cr
&+ &\sum_{s=1}^{r_1-1}\left(\sum_{l=0}^{R_2}
  \mu^s Z_l^{(s)}\Big(\rho,\frac{\csi^2+\eta^2}{2},\lambda\Big)
+ \sum_{l>R_2}\mu^{s}f_l^{(r_1,r_2-1;s)}(\rho,\csi,\lambda,\eta)\right)
\cr
&+ &\sum_{l=0}^{r_2-1}\mu^{r_1}
 Z_l^{(r_1)}\Big(\rho,\frac{\csi^2+\eta^2}{2},\lambda\Big)
\,+\,\sum_{l\ge r_2}\mu^{r_1}f_l^{(r_1,r_2-1;r_1)}(\rho,\csi,\lambda,\eta)
\cr
&+ &\sum_{s>r_1}\sum_{l\ge 0}\mu^sf_l^{(r_1,r_2-1;s)}(\rho,\csi,\lambda,\eta)\ ,
\cr
}}
\label{eq:H(r1,r2-1)}
\end{equation}
where $Z_l^{(0)}\in\Pscr_{l,0}$ $\forall\ l\ge 4$,
$Z_l^{(s)}\in\Pscr_{l,s}$ $\forall\ 0\le l\le R_2\,,\ 1\le s<r_1\,$,
$Z_l^{(r_1)}\in\Pscr_{l,r_1}$ $\forall\ 0\le l<r_2\,$,
$f_l^{(r_1,r_2-1;r_1)}\in\Pscr_{l,r_1}$ $\forall\ l\ge r_2\,$,
$f_l^{(r_1,r_2-1;s)}\in\Pscr_{l,s}$ $\forall\ l>R_2\,,\ 1\le s<r_1\,$
and $\forall\ l\ge 0,\ s>r_1\,$. Let us recall that the Hamiltonian
$H^{(1,0)}=\Hscr$ written in~\eqref{eq:initial-Ham-rewritten} is
suitable for starting the procedure with $r_1=r_2=1$.  In
formula~\eqref{eq:H(r1,r2-1)}, one can distinguish the normal form
terms from the perturbing part; the latter depends on $(\csi,\eta)$ in
a {\it generic} way, while in the terms of $Z$ type appearing
in~\eqref{eq:H(r1,r2-1)} the fast variables can be replaced by the
action $\Gamma=(\csi^2+\eta^2)/2\,$. The $(r_1,r_2)$--th step of the
algorithm formally defines the new Hamiltonian $H^{(r_1,r_2)}$ by
applying the Lie series operator
$\exp\Lie_{\mu^{r_1}\chi_{r_2}^{(r_1)}}$ to the previous Hamiltonian
$H^{(r_1,r_2-1)}$, as it is prescribed in
formula~\eqref{eq:def-funzionale-H(r1,r2)}.  The new generating
function $\mu^{r_1}\chi_{r_2}^{(r_1)}$ is determined by solving the
following homological equation with respect to the unknown
$\chi_{r_2}^{(r_1)}=\chi_{r_2}^{(r_1)}(\rho,\csi,\lambda,\eta)$:
\begin{equation}
\Lie_{\chi_{r_2}^{(r_1)}}Z_2^{(0)}+f_{r_2}^{(r_1,r_2-1;r_1)}=Z_{r_2}^{(r_1)}\ ,
\label{eq:chi(r1,r2-1)}
\end{equation}
where we require that $Z_{r_2}^{(r_1)}$ is the new term in normal
form, i.e.
$Z_{r_2}^{(r_1)}=Z_{r_2}^{(r_1)}\big(\rho,(\csi^2+\eta^2)/2,\lambda\big)$.

\begin{proposition}
If $Z_2^{(0)}=(\csi^2+\eta^2)/2$ and
$f_{r_2}^{(r_1,r_2-1;r_1)}\in\Pscr_{r_2,r_1}\,$, then there exists a
generating function $\chi_{r_2}^{(r_1)}\in\Pscr_{r_2,r_1}$ and a
normal form term $Z_{r_2}^{(r_1)}\in\Pscr_{r_2,r_1}$  satisfying
  the homological equation~\eqref{eq:chi(r1,r2-1)}.
\label{lem:sol_homol_eq}
\end{proposition}
We limit ourselves to just sketch the procedure that can be followed
so as to explicitly determine a solution of~\eqref{eq:chi(r1,r2-1)}
and, therefore, prove the statement above. First, we replace the fast
coordinates $(\csi,\eta)$ with the pair of complex conjugate canonical
variables $(z,\imunit{\overline z})$ such that $\csi=(z-{\overline
  z})/\sqrt{2}$ and $\eta=(z+{\overline z})/\sqrt{2}$. Moreover, the
homological equation~\eqref{eq:chi(r1,r2-1)} has to be expanded in
Taylor series with respect to $(z,\imunit{\overline z})$, using the
slow coordinates $(\rho,\lambda)$ as fixed parameters (because they
are not affected by the Poisson bracket
$\Lie_{\chi_{r_2}^{(r_1)}}Z_2^{(0)}$, since $Z_2^{(0)}$ do not depend
on them). Therefore, we solve term-by-term the
equation~\eqref{eq:chi(r1,r2-1)} in the unknown
coefficients\footnote{We emphasize that the most celebrated problem
  concerning the convergence of the normal forms, i.e., the occurrence
  of ``small divisors'', does not affect our scheme, because the main
  integrable term in the homological equation~\eqref{eq:chi(r1,r2-1)},
  i.e. $Z_2^{(0)}$, depends just on the fast action
  $(\csi^2+\eta^2)/2$.}  $x_{m_1,m_2,m_3,k_1,k_2,j}$ and
$\zeta_{m_1,m_2,m_2,k_1,k_2,j}$ such that
$$
\vcenter{\openup1\jot\halign{
 \hbox {\hfil $\displaystyle {#}$}
&\hbox {\hfil $\displaystyle {#}$\hfil}
&\hbox {$\displaystyle {#}$\hfil}\cr
\chi_{r_2}^{(r_1)}(\rho,z,\lambda,\imunit{\overline z}) &=
&
\sum_{\scriptstyle{2m_1+m_2}\atop{\scriptstyle {+m_3=l}}}
\ \sum_{{\scriptstyle{k_1+k_2\le l+4r_1-3}}\atop{\scriptstyle {j\le 2l+7r_1-6}}}
x_{m_1,m_2,m_3,k_1,k_2,j}\,\rho^{m_1}z^{m_2}(\imunit{\overline z})^{m_3}
\,(\cos\lambda)^{k_1}(\sin\lambda)^{k_2}\,\big(\beta(\lambda)\big)^j
\cr
}}
$$
and
$$
\vcenter{\openup1\jot\halign{
 \hbox {\hfil $\displaystyle {#}$}
&\hbox {\hfil $\displaystyle {#}$\hfil}
&\hbox {$\displaystyle {#}$\hfil}\cr
Z_{r_2}^{(r_1)}(\rho,z,\lambda,\imunit{\overline z}) &=
&
\sum_{\scriptstyle{2m_1+{\phantom{l}}}\atop{\scriptstyle {2m_2=l}}}
\ \sum_{{\scriptstyle{k_1+k_2\le l+4r_1-3}}\atop{\scriptstyle {j\le 2l+7r_1-6}}}
\zeta_{m_1,m_2,m_2,k_1,k_2,j}\,\rho^{m_1}(z\cdot\imunit{\overline z})^{m_2}
\,(\cos\lambda)^{k_1}(\sin\lambda)^{k_2}\,\big(\beta(\lambda)\big)^j\ .
\cr
}}
$$ 
At last, we express the expansions above by replacing
$(z,\imunit{\overline z})$ with $(\csi,\eta)$, and we obtain the final
solutions in the form
$\chi_{r_2}^{(r_1)}=\chi_{r_2}^{(r_1)}(\rho,\csi,\lambda,\eta)$ and
$Z_{r_2}^{(r_1)}=Z_{r_2}^{(r_1)}\big(\rho,(\csi^2+\eta^2)/2,\lambda\big)$.

The following property of the Poisson brackets is very useful for our
purposes; however, since its proof just requires long but
basically trivial calculations, it is omitted.

\begin{proposition}
Let $f$ and $g$ be two generic functions such that $f\in\Pscr_{r,s}$
and $g\in\Pscr_{r^{\prime},s^{\prime}}\,$, then
$$
{\rm if}\ r+r^{\prime}\ge 2\ \ \Rightarrow
\ \ \poisson{f}{g}\in\Pscr_{r+r^{\prime}-2,s+s^{\prime}}\ ,
\qquad
{\rm else}\ \ \Rightarrow\ \ \poisson{f}{g}=0\ .
$$
\label{lem:alg-prop-Pois-brackets}
\end{proposition}

\noindent 
In order to provide an algorithm easy to translate in a programming
language, we are going to give explicit formulas for the new terms of
type $f$ appearing in expansion~\eqref{eq:H(r1,r2)} of Hamiltonian
$H^{(r_1,r_2)}$.  Let us introduce the recursive operation
$a\pluseq b$, where the previously defined quantity $a$ is redefined
as $a=a+b\,$. Therefore, we initially define
\begin{align}
f_l^{(r_1,r_2;s)} = f_l^{(r_1,r_2-1;s)} \qquad & \forall\ l>R_2 \ {\rm when}\ 1\le s<r_1 \ {\tt or} \nonumber\\
& \forall\ l>r_2\ {\rm when}\ s=r_1
\ {\tt or}\  \forall\ l\ge 0\,,\ s> r_1~~. \nonumber\\
\nonumber
\end{align}
Then, we consider the contribution of the terms generated by the Lie
series applied to each function belonging to the normal form part as
follows:
\begin{equation}
f_{l+j(r_2-2)}^{(r_1,r_2;s+jr_1)}\pluseq
\frac{1}{j!}\Lie_{\chi_{r_2}^{(r_1)}}^jZ_l^{(s)}
\qquad\ \forall\ 1\le j< {\bar j}_f\,,
\ 0\le l< {\bar l}_f\,,\ 0\le s\le r_1\ ,
\label{eq:f_l^r1r2s_def_2}
\end{equation}
where the upper limits ${\bar j}_f$ and ${\bar l}_f$ on the indexes
$j$ and $l$, respectively, are such that
\begin{equation}
\vcenter{\openup1\jot\halign{
 \hbox {\hfil $\displaystyle {#}$}
&\hbox {$\displaystyle {#}$\hfil}
\ \quad
&\hbox {\hfil $\displaystyle {#}$}
&\hbox {$\displaystyle {#}$\hfil}
\ \quad
&\hbox {\hfil $\displaystyle {#}$}
&\hbox {$\displaystyle {#}$\hfil}
\cr
{\bar j}_f=l+1 &\ {\rm if}\ r_2=1\ ,
&{\bar j}_f=+\infty &\ {\rm if}\ r_2\ge 2\ ,\cr
{\bar l}_f=+\infty &\ {\rm if}\ s=0\ ,
&{\bar l}_f=R_2+1 &\ {\rm if}\ 1\le s<r_1\ ,
&{\bar l}_f=r_2 &\ {\rm if}\ s=r_1\ ,\cr
{\bar l}_i=R_2+1 &\ {\rm if}\ 1\le s<r_1\ ,
&{\bar l}_i=r_2 &\ {\rm if}\ s=r_1\ ,
&{\bar l}_i=0 &\ {\rm if}\ s>r_1\ .\cr
}}
\label{eq:limits-indexes-j-l}
\end{equation}
For what concerns the contributions given by the perturbing terms
making part of the expansion of $H^{(r_1,r_2-1)}$
in~\eqref{eq:H(r1,r2-1)}, we have
\begin{equation}
f_{l+j(r_2-2)}^{(r_1,r_2;s+jr_1)}\pluseq
\frac{1}{j!}\Lie_{\chi_{r_2}^{(r_1)}}^jf_l^{(r_1,r_2-1;s)}
\qquad\ \forall\ 1\le j< {\bar j}_f\,,
\ l\ge {\bar l}_i\,,\ s\ge 1\ ,
\label{eq:f_l^r1r2s_def_3}
\end{equation}
where the limiting values for the indexes, that are ${\bar j}_f$ and
${\bar l}_i\,$, are defined in~\eqref{eq:limits-indexes-j-l}.

The redefinition rules~\eqref{eq:f_l^r1r2s_def_2}
and~\eqref{eq:f_l^r1r2s_def_3} are set so that the new perturbing part
generated by the Lie series in~\eqref{eq:def-funzionale-H(r1,r2)} is
coherently split in different terms according to their order of
magnitude in $\mu$ and their total polynomial degree in the
actions. In fact, by applying repeatedly
proposition~\ref{lem:alg-prop-Pois-brackets} to the redefinitions
in~\eqref{eq:f_l^r1r2s_def_1}--\eqref{eq:f_l^r1r2s_def_3}, it is
possible to inductively verify that $f_l^{(r_1,r_2;s)}\in\Pscr_{l,s}$
$\forall\ l\ge {\bar l}_i,\ s\ge 1\,$. Therefore, the terms making
part of the Hamiltonian~$H^{(r_1,r_2)}$ in the
expansion~\eqref{eq:H(r1,r2)} share the same properties with those
appearing in~\eqref{eq:H(r1,r2-1)}; this ensures that the
normalization algorithm can be iterated so as to construct
$H^{(r_1,r_2+1)},\ H^{(r_1,r_2+2)},\ \ldots\,,\ H^{(r_1,R_2)}$.  The
simple prescription $H^{(r_1+1,0)}=H^{(r_1,R_2)}$, $\forall\ 1\le
r_1<R_1\,$, allows to deal with generating functions of the next order
of magnitude, i.e., $\Oscr(\mu^{r_1+1})$. As discussed in
subsection~\ref{sbs:average}, by applying repeatedly the rules of the
generic normalization step, the algorithm finally ends, by determining
the last Hamiltonian $H^{(R_1,R_2)}$.


\end{document}